\documentclass[aps,prl,twocolumn,showpacs,superscriptaddress,groupedaddress,doublespaced]{revtex4}
\usepackage{amsmath,amssymb}
\usepackage{graphicx}% Include figure files
\usepackage{dcolumn}% Align table columns on decimal point
\usepackage{bm}% bold math
\usepackage{subfig}
\usepackage{color}
\newcommand{\ber}{\begin{eqnarray}}
\newcommand{\eer}{\end{eqnarray}}
\newcommand{\bea}{\begin{equation}}
\newcommand{\eea}{\end{equation}}

\begin{document}

\title{Random walk model for coordinate dependent diffusion in a force field}
\author{Rohan Maniar}
\author{A. Bhattacharyay}
\email{a.bhattacharyay@iiserpune.ac.in}
\affiliation{Indian Institute of Science Education and Research, Pune, India}

\begin{abstract}
In this paper we develop a random walk model on a lattice for coordinate dependent diffusion at constant temperature. We employ here a coordinate dependent waiting time of the random walker to make the diffusivity coordinate dependent. The presence of a confining conservative force is modeled by appropriately breaking the isotropy of the jumps of the random walker to its nearest neighbors. We show that the equilibrium is characterized by the position distribution which is of modified Boltzmann form. We also show that, in such systems with coordinate dependent diffusivity, the modified Boltzmann distribution correctly captures the transition over a potential barrier as opposed to the Boltzmann distribution.
\end{abstract}
%Use showkeys class option if keyword
\maketitle

\section{Introduction}
\vspace{3mm}
Coordinate dependent diffusion is ubiquitous. It is believed to play a significant role in the folding transitions of bio-molecules like proteins. Shift of kinetic transition and barrier height of protein folding based on coordinate dependent diffusion has been demonstrated by Chahine at al. \cite{chah}. There are important investigations on the role of position-dependent diffusion in protein (bio-molecule) folding in the ref.\cite{best,bere,fost}. The role of position dependent diffusion in oxygen transport through lipid membrane has been explored by Ghysels et al. in \cite{ghys}. In the context of fabrication of multijunction solar cells, coordinate-dependent diffusion of phosphorus in gallium doped germanium has been investigated \cite{kobe}. Direct experimental evidence of position dependent diffusion of light in disordered waveguides has been reported by Yamilov et al. \cite{yami}
\\

Coordinate dependent diffusion of a Brownian particle (BP) when modeled by Langevin dynamics causes controversy because it involves multiplicative noise \cite{soko,tupp,leib}. The source of controversy is in the stochastic integrals which result in a diffusion gradient driven spurious drift current in the corresponding Fokker-Plank equation. The controversy is mainly in physics literature where Stratonovich or Stratonovich-like conventions are used for stochastic integrations involving multiplicative noise \cite{lau,san1,san2,mark,far1,far2}. These conventions produce spurious currents proportional to the diffusivity gradients which are then removed by adding counter terms in the Langevin dynamics. The removal of a part of the spurious probability current from the Fokker Plank dynamics is done predominantly in the existing physics literature by demanding the Boltzmann distribution (BD) to ensure equilibrium in such systems \cite{lau}. It has been pointed out by one of the authors of this paper earlier that while removing the spurious part of the probability current one also tacitly ignores a part of legitimate diffusion current in order to have a Boltzmann distribution \cite{ari1,ari2}. Otherwise, the Boltzmann Distribution will not come out to be the equilibrium distribution of such a system if the diffusion current is kept as it should be for coordinate dependent diffusion.
\\

The motivation of the present paper is a numerical exploration of such a system of BP using a model which does not involve controversy like the one related to the existence of multiplicative noise in the Langevin dynamics. We investigate here the equilibrium and weakly non-equilibrium escape rates within the scope of Kramers' escape rate theory of a BP with coordinate dependent diffusivity. Employing Langevin dynamics to such a system is controversial due to the standard dilemma of adding or not adding additional terms to remove spurious currents. On the other hand molecular dynamics simulations of more realistic systems are much more computationally expensive. The question of employing Monte Carlo simulations does not exist when the controversy on the distribution of the system is being addressed. 
\\

In the present paper we first use the random walk based numerical investigation to show that the equilibrium distribution of a confined Brownian particle with coordinate dependent diffusivity is indeed a modified Boltzmann distribution. The present method of modeling with random walk is simple, elegant and could prove to be extremely useful for simulation of coordinate dependent diffusion in a conservative force field. The outcome of the simulation is extremely important in establishing the modified Boltzmann distribution in the presence of coordinate dependent diffusivity and this is fully consistent with the previous analytic predictions of one of the present authors \cite{ari1,ari2}.
\\

Once the numerical method is thus established, we use the same to probe transition rate over a potential barrier in the presence of coordinate dependent diffusivity. We show that, within the purview of Kramers' rate theory, where the Boltzmann distribution fails to match simulation results, the modified Boltzmann distribution captures the process quite accurately and this is the main result of the present paper. It turns out that, in the presence of coordinate dependent diffusivity, the diffusivity near the barrier dominates transition rate when Boltzmann distribution is employed. Whereas, for the modified Boltzmann distribution, the diffusivity inside the potential well minimum is the key factor. Now Kramers' escape rate process of transition over an energy barrier being dominated by the equilibrium distribution of a BP in the potential well the diffusivity near the barrier cannot clearly dominate the transitions because the particle spend very little time at the barrier. This clear distinction between the role of these two distributions could be crucial for experimental determination of which one is at work in such systems.
\\

Using position dependent hopping probabilities of a Brownian walker on a lattice is not new \cite{gate, codl, greb, hou, long}. This model has been extensively used in investigating heterogeneous diffusion processes when the coordinate dependent variation of hopping are in general drawn from some distribution to model a heterogeneous environment. In general, the bias in the jumping probability resulting from anisotropy can model the  presence of a force on an over-damped particle and this method can be quite general \cite{falc}. This is exactly what is utilized in the present work to represent a confining force to a random walker in a one dimensional lattice for the sake of simplicity.
\\

The other important element which has been made use of is the waiting time of such a walker at a lattice site to generate coordinate dependent diffusion at a constant temperature. By introducing a waiting time globally at each lattice point one can control the diffusivity of the walker globally. A local generalization of diffusivity then requires some coordinate dependence of this waiting time. The most important feature of this way of introducing diffusivity variation, as will be shown in this paper, is that it can be done over a constant temperature environment under a confinement. Once standardized, this very simplistic random walk model can actually capture Brownian motion over an inhomogeneous space when the inhomogeneity is not only due to the presence of a conservative force but also due to varied diffusivity over space. 
\\

In this paper we first establish the usefulness of this (above mentioned) method of variation of diffusivity over space using a BP under harmonic confinement. Following which we embark on simulating coordinate dependent Brownian motion by employing a waiting time gradient over space. We obtain the equilibrium distribution of such a walker to show that, besides having the Boltzmann factor, the distribution has an amplitude which is inversely proportional the coordinate dependent diffusivity. After this, we present a weakly non-equilibrium process of Brownian particles overcoming a barrier and establish the modified Boltzmann distribution \cite{ari1,ari2} through this Kramers' escape rate process.

\section{Random Walk Model of Langevin Dynamics in homogeneous space}
\vspace{3mm}
The Langevin dynamics of the form
\\
\begin{equation}
\Gamma\dot{x}=F(x)+\Gamma\sqrt{2D}\eta(t),
\end{equation}
\\
where $x$ is the position of the BP in a force field $F(x)=-\frac{\partial V(x)}{\partial x}$ and $\Gamma$ is a damping constant, can be modeled by a random walker. In the above equation $D$ is the diffusivity of the particle and $\eta(t)$ is a Gaussian white noise of unit strength. As has been mentioned in the introduction the purpose of modeling this Langevin dynamics by random walks is to extend this model to a situation where the $\Gamma$ and $D$ are coordinate dependent. However, in the beginning we are interested in establishing our random walk model for uniform diffusivity in order to demonstrate consistency of the method in tuning the diffusivity.
\\

Consider the force free situation when Eq.(1) takes the shape $\frac{\partial x}{\partial t}=\sqrt{2D}\eta(t)$ being represented by a random walker on a one dimensional lattice with jumping probability towards the nearest neighbour points being $\frac{1}{2}$. In such a situation, one would get a pure diffusion with highest possible diffusivity on this lattice compared to the process when the particle has a non-zero probability $p_{s}$ to stick to the lattice point. In the presence of a probability to stick to the lattice point being $p_{s}$ and the jumping probabilities to the right and left being $\frac{1-p_{s}}{2}$ the diffusivity of the particle will fall from this highest value. We alternately define the moving probability $p_{m}$ which is defined as $1-p_{s}$.
\\
 
Fig.1 shows such a plot of a random walker's diffusivity as a function of $p_{m}$ where the diffusivity has been obtained by evaluating numerically $D=\langle \frac{(x(t)-x(0))^{2}}{2t}\rangle$ for a long enough time t and $\langle *\rangle$ indicates an ensemble average over $10^5$ particles. Everywhere in what follows (if not mentioned otherwise) all the ensemble averages are over this set of $10^5$ particles. Corresponding to all the points in the graph (Fig.1) there is a constant $p_{s}$ everywhere on the lattice characterizing a constant global diffusivity. The time is measured in unit steps in which the particle either jumps to the right or left nearest lattice points with probability $\frac{1-p_{s}}{2}$ to maintain isotropy of the process or sticks to the point with a probability $p_{s}$. The dependence as indicated by Fig.1 is linear with a $y$-intercept of zero and on performing a linear fit the slope of 0.497 is found. To understand this graph the definition of the diffusivity helps. This means that, for a constant MSD ($\langle {(x(t)-x(0))^{2}}\rangle=1$) the unit of time interval for different $p_m$ is effectively scaling as $1/p_m$. For a larger $p_m$ the particle makes a jump over an effectively shorter time interval and for smaller $p_m$ it is the opposite. Thus, for a unit MSD i.e. $\langle {(x(t)-x(0))^{2}}\rangle =1$, one gets the linear dependence of the diffusivity $D$ on $p_m$ and that gives a slope of half (1/2) for the straight line. Thus the slope of half reflects the dimension of the space. Having known the dependence of the diffusivity on this waiting time probability $p_{s}$, we can make $D$ a function of space (coordinates) by making $p_{s}$ a function of the coordinates.
\\ 
\\
\begin{figure}[h!]
  \includegraphics[width=60mm]{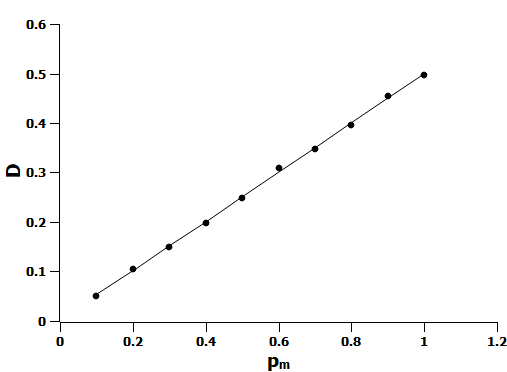}
   \caption{The plot for Diffusivity as a function of moving probability.}
\end{figure}
\\

To understand how a force can be brought in this random walk model note that, in the presence of a force $F(x)$, since there will be a drift velocity $v_{d}=\frac{F(x)}{\Gamma}$, one can accommodate the presence of this force on the Brownian motion by breaking the isotropy of the jump probabilities towards and against the force in the following ways. The jump probability towards the force is $p_{t}=\frac{1-p_{s}}{2}+\epsilon\frac{F(x)}{\Gamma}$ and against the force is $p_{a}=\frac{1-p_{s}}{2} -\epsilon\frac{F(x)}{\Gamma}$ where $\epsilon$ is a dimension fixing constant. Note that, the $\Gamma$ in the Langevin dynamics (Eq.(1)) is just a scale factor in units of which we are going to get the force term for a known diffusivity $D$. The factor $\epsilon/\Gamma$ is in general set to be a small number to determine how far up the potential $V(x)$ the BP will rise under given conditions of the force.
\\

We can simplify the terms in $p_{a}$ and $p_{t}$ to get a more illuminating form for these jump probabilities by evoking the Stoke Einstein relation and using the relation $D=0.497\times p_{m}$ shown to hold previously.
\\
\\
$p_{a}=\frac{1-p_{s}}{2} - \epsilon\frac{F(x)}{\Gamma} = \frac{p_{m}}{2} - \epsilon\frac{F(x)D}{k_{B} T}=\frac{p_{m}}{2} - \epsilon\frac{0.497 F(x) p_{m}}{k_{B}T}$
\\
$= p_{m} \Big[\frac{1}{2} -\epsilon\frac{0.497 F(x)}{k_{B}T} \Big] \simeq p_{m} \Big[\frac{1}{2} -\frac{F(x)}{2N_{0}} \Big]$
\\
\\
where over here we have defined $\frac{1}{N_{0}}\simeq\frac{\epsilon}{k_{B}T}$. Thus, the temperature of the system can be controlled using $N_{0}$. Now, to keep the total drift velocity $v_d = v_+ -v_-=F(x)/\Gamma$ where $v_\pm$ are the velocities in the direction of and against the force, we easily identify that $\epsilon = 1/2$ in the present model. Note that, while the above simplification was done for the case of a constant diffusivity it can easily be extended to the case of a coordinate dependent diffusivity where the moving probability $p_{m}$ will be coordinate dependent. 
\\
\begin{figure}
\subfloat[$p_{s}=0.2$]{\includegraphics[width=60mm]{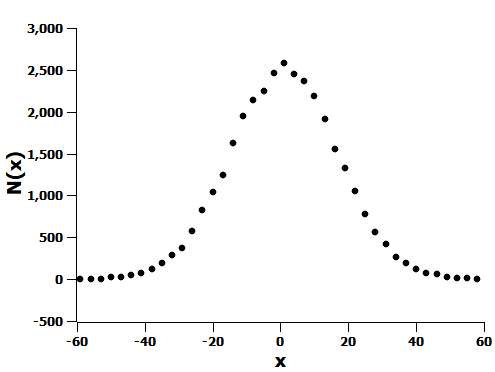}}\\
\subfloat[$p_{s}=0.6$]{\includegraphics[width=60mm]{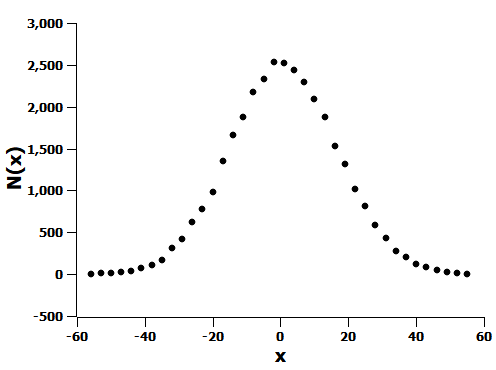}}
\caption{Gaussian distribution for $p_{s}=0.2$ (Fig.2(a)) and $ p_{s}=0.6$ (Fig.2(b))}
\end{figure}

As a simple example, consider a harmonic potential $V(x)=\frac{kx^2}{2}$ where the drift velocity is $\frac{-kx}{\Gamma}$. For simplicity we will set $k=1$ in what follows. By setting $p_{s}=0.2$ and $p_{s}=0.6$ with $N_{0}=500$ in the presence of only nearest neighbour jumps, we evolve the walker and arrive at the Gaussian distributions of the position of the walker as shown in Fig.2(a) and (b). To verify the Gaussian nature of the figures mentioned we plot the log of number of particles at a position coordinate x vs $x^{2}$ and performed a linear fit. This was done for x between $-60$ and $60$ and the R square for this fit was $0.9920$.
\\

We can perform a few immediate checks here. We find out the diffusivity of the systems corresponding to any $p_s$ value. Having known that the mean square displacement (MSD) of the BP in the presence of the force will go as
\begin{equation}
<x^2(t)>=\frac{D\Gamma}{k}(1-e^{\frac{-2kt}{\Gamma}})=A(1-e^{-Rt})
\end{equation}
we get the universal relation $\frac{RA}{D}=2$. Evidence for the same is presented in Table.1 where we show that the universal relation holding for different values of $p_{s}$. Fig.3(a) shows the time variation of the MSD for $p_s=0.2$, $0.4$ and $0.6$ where $N_0 = 600$ and Fig.3(b) shows the plot of MSD against time for $p_s=0.4$ and $N_0=200$, $400$, $600$. Note the important feature shown by these two graphs that the saturation value of the MSD being $A=\frac{D\Gamma}{k}$ varies with $N_0$ and not with $p_s$. Thus, the temperature of the system $\frac{D\Gamma}{k}=\frac{k_BT}{k}$ in equilibrium where $D\Gamma =k_BT$ with $k_B$ being Boltzmann constant could be tuned by the $N_0$.
\\

One can also verify from Fig.3(b) that the relation $ \frac{1}{N_{0}} = \frac{1}{2k_{b}T} $ holds by noting that for the case of $k=1$ the equilibrium distribution is given by $P(x)=N exp(-\frac{x^{2}}{2k_{b}T})$. This would imply that the variance of the gaussian particle distribution, given by the saturation value of $<x^{2}(t)>$ in the figure, is equal to $k_{b}T$ which agrees well with the claimed relation. The other important thing is that, the variation of the diffusivity by the change of the sticking probability does not affect the temperature of the system as is shown by Fig.3(a). This particular feature of the tuning of the diffusivity by $p_s$ will allow us to introduce a coordinate dependent diffusivity by making $p_s$ a function of space without changing the temperature. 

\begin{figure}
\subfloat[Time dependence of MSD for $p_{s}=0.2$,$0.4$,$0.6$ where $N_{0}=600$]{\includegraphics[width=60mm]{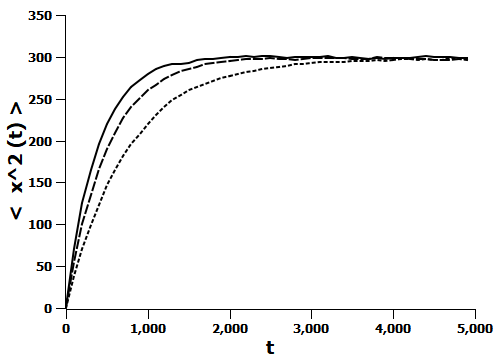}}\\
\subfloat[Time dependence of MSD for $p_{s}=0.4$ where $N_{0}=200$,$400$,$600$]{\includegraphics[width=60mm]{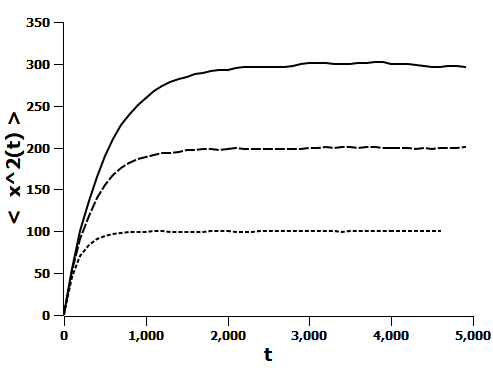}}
\caption{Time dependence of MSD for a BP in a harmonic potential for varying values of $p_{s}$ and $N_{0}$}
\end{figure}
\hspace*{4mm}

\begin{table}
\caption{Verification of the universal relation $\frac{RA}{D}=2$ for different values of $p_{S}$ where $N_{0}=500$}
\begin{center}
 \begin{tabular}{|c c c c c|} 
 \hline
 $p_{s}$ & R & A & D & $\frac{RA}{D}$ \\ [0.5ex] 
 \hline\hline
 0 & 0.00389 & 253.77 & 0.506 & 1.95 \\ 
% \hline
 0.1 & 0.00365 & 249.18 & 0.447 & 2.05 \\
% \hline
 0.2 & 0.00323 & 250.37 & 0.396 & 2.03 \\
% \hline
 0.3 & 0.00277 & 251.42 & 0.354 & 1.97 \\
% \hline
 0.4 & 0.00250 & 244.27 & 0.305 &1.99 \\  
% \hline
 0.5 & 0.00197 & 251.66 & 0.249 & 1.99\\
% \hline
 0.6 & 0.00161 & 247.68 & 0.199 & 2.01\\
% \hline
 0.7 & 0.00120 & 249.01 & 0.150 & 1.99\\
% \hline
 0.8 & 0.00080 & 252.20 & 0.099 & 2.05\\[0.5ex]
 \hline
\end{tabular}
\end{center}
\end{table}

\section{Random Walk Model of Langevin Dynamics with coordinate dependent diffusivity}

Let us first employ a ramp in the diffusivity of the above mentioned model system by introducing a gradient of the sticking probability $p_s$. This is done by varying the sticking probability as follows:

\[p_{s}=\begin{cases}
		a &x\leq-25\\
		\frac{(b-a)(x-25)}{50}+b &-25\leq x \leq25\\
		b &25\leq x\\
		\end{cases}
\]
where a and b are any two values of $p_{s}<1$. In Fig.4(a) and (b) we show the modified stationary Gaussian distribution of the system for $a=0.4$, $b=0.8$, $N_{0}=1000$ and $a=0.2$, $b=0.8$, $N_{0}=500$ respectively. The numerically obtained distributions are fitted to the distributions $P(x)=\frac{N}{D(x)}\exp{(-kx^2/2D\Gamma)}=\frac{N}{D(x)}\exp{(-kx^2/2k_BT)}$ and we obtain the parameter $k_BT/k$ from the MSD of the system as is set by the fixed $N_0$ and the $D(x)$ is found out from its correspondence with the $p_s$ variation as shown in Fig.1. In the above mentioned expression of modified Boltzmann distribution $N$ is a normalization constant.
\\

To verify that the stationary distribution does indeed follow the modified BD we plot log of number of particles at $x$ multiplied with the value of the diffusivity at $x$ vs $x^{2}$ and perform a linear fit. The linear fits for the first and the second systems are shown in Fig.4(c) and Fig.4(d) respectively. The fit is done taking x between $-42$ and $42$ and the R square values for these fits are $0.985$ and $0.993$ respectively. The excellent agreement of the numerical distribution with the theoretical one demonstrate the existence of the modified Boltzmann distribution of such a BP under confinement in equilibrium with a heat bath.
\begin{figure}
\subfloat[Stationary distribution for the system $a=0.4$,$b=0.8$,$N_{0}=1000$]{\includegraphics[width= 60mm]{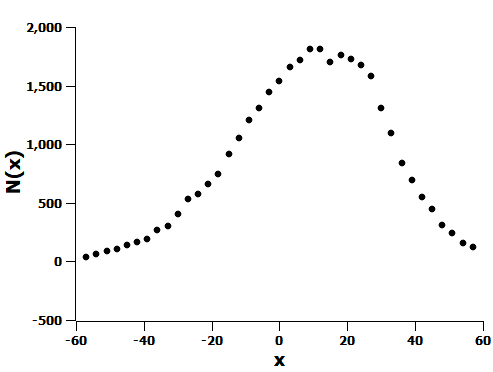}}\\
\subfloat[Stationary distribution for the system $a=0.2$,$b=0.8$,$N_{0}=500$]{\includegraphics[width= 60mm]{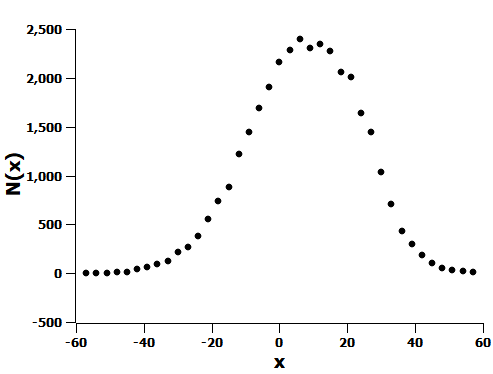}}\\
\subfloat[Linear fit for the system $a=0.4$,$b=0.8$,$N_{0}=1000$]{\includegraphics[width= 60mm]{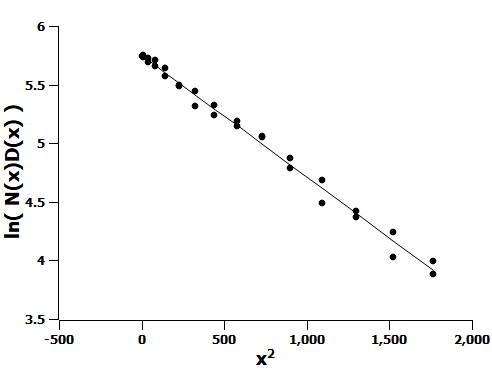}}\\
\subfloat[Linear fit for the system $a=0.2$,$b=0.8$,$N_{0}=500$]{\includegraphics[width= 60mm]{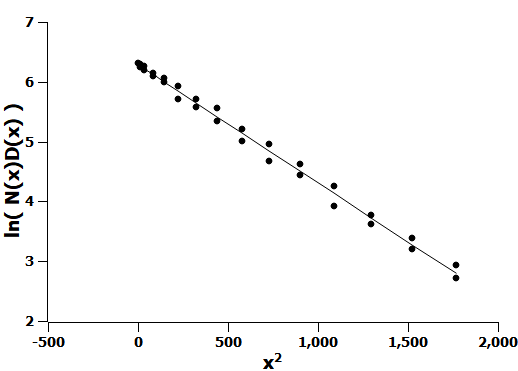}}
\caption{Modified BD for the case of coordinate dependent diffusivity in a harmonic potential}
\end{figure}
\\

We generalize this method to the case of a double well potential to show that the present model can indeed be utilized for more structured confinements. The double well potential is of the form $V(x)=-\alpha x^2 + \beta x^4$ using the same procedure as mentioned for the harmonic confinement. The simulation of the double well system $N_0 =2000$ with $\alpha = 5.0$ and $\beta = 0.0125$ are set. Fig.5(a) shows the stationary distribution for particles in the case of a double well potential with a coordinate dependent $p_{s}$ ramp. The sticking probability was varied as

\[p_{s}=\begin{cases}
		0.2 &x\leq-100\\
		\frac{(0.6)(x-100)}{200}+0.8 &-100\leq x \leq100\\
		0.8 &100\leq x .\\
		\end{cases}
\]
\\

Fig.5(b) is a plot of log of number of particles at x multiplied with the diffusivity at x vs x. The curve is fit to a function of the form $f(x)= A + cx^{2} -dx^{4}$. Here, the shift $A = \ln{N}$ where $N$ is the overall normalization constant. From the fit, the magnitude of the constants $c$ and $d$ are coming out to be $c=0.0051$ and $d=1.272\times 10^{-5}$. This shows that the ratio of $\alpha$ to $\beta$ is almost the same as that of $c$ to $d$ which it should be. This was done for x between -27 and 27 and the R square value of the fit was $0.996$. For comparison, Fig.5(c) is a plot of log of number of particles at x vs x where the continuous curve is the same function $f(x)= A+cx^{2}-dx^{4}$ . The left right asymmetry clearly indicates the absence of the $D(x)$ dependent amplitude of the distribution which is missing in this case.
\\
\begin{figure}
\subfloat[Stationary distribution for the double well potential with a $p_{s}$ ramp]{\includegraphics[width= 60mm]{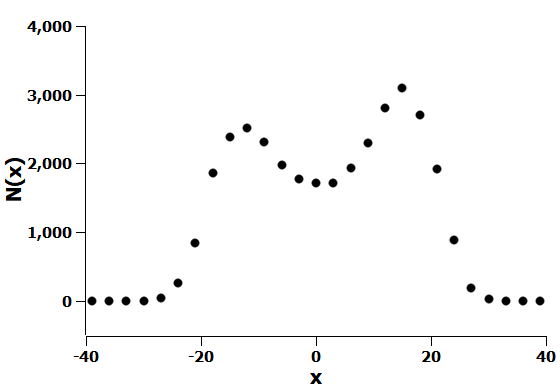}}\\
\subfloat[Fit curve for the modified BD]{\includegraphics[width= 60mm]{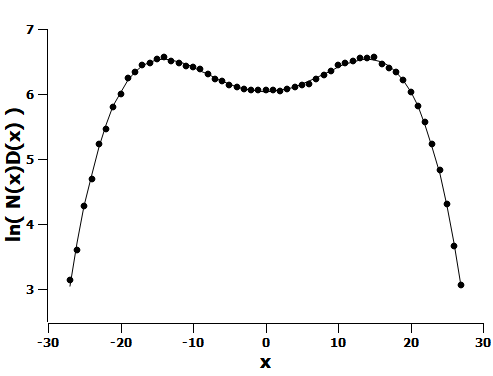}}\\
\subfloat[Fit curve for the BD]{\includegraphics[width= 60mm]{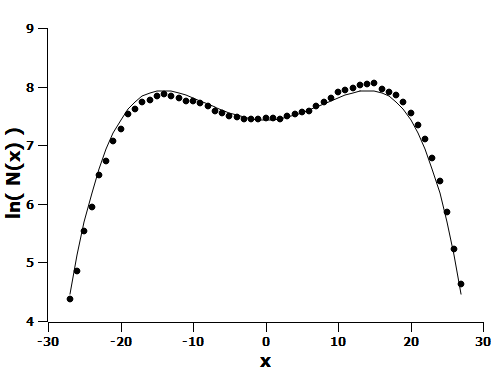}}
\caption{Modified BD for the case of coordinate dependent diffusivity in a double well potential}
\end{figure}

Having seen the numerical results of the equilibrium distribution of a BP with coordinate dependent diffusivity and damping let us have a look at the theoretical derivation of the same for the sake of completeness. In the following we show the derivation of the equilibrium distribution of the BP following the method shown in \cite{ari1,ari2} which relies on Smoluchowski equation obtained by adopting It\^o convention for multiplicative noise.
\\

Consider the overdamped dynamics of the BP when its diffusivity and damping are functions of coordinate.
\bea
\dot{x} = \frac{F(x)}{\Gamma(x)} + \sqrt{2{D(x)}}\eta(t).
\eea
Here, $D(x)$ is a coordinate dependent diffusivity in 1D and $\Gamma(x)$ is the coordinate dependent damping. The corresponding Smoluchowski equation in It\^o convention will look like
\bea
\frac{\partial {P(x,t)}}{\partial t} = \frac{\partial}{\partial x}\left [-\frac{{F(x)}{P(x,t)}}{\Gamma(x)} + \frac{\partial D(x){ P(x,t)}}{\partial x} \right ].
\eea
\\

The $P(x,t)$ in the above equation is the probability density of the BP. For equilibrium distribution of this probability density, we have to consider it to be independent of time and the distribution will follow from the condition of detailed balance which requires the square bracketed terms on the right hand side of the above equation set to zero. The detailed balance will straight forwardly give the equilibrium distribution of the system to be
\bea
\text P(x) = \frac{\text N}{ D(x)}\exp{\int_{-\infty}^x \frac{ F(x^\prime)}{ D(x^\prime)\Gamma(x^\prime)} dx^\prime},
\eea
where N is a normalization constant. This distribution can be written as 
\bea
\text P(x) = \frac{\text N}{ D(x)}\exp{\frac{-V(x)}{ k_BT}},
\eea
when the Stokes-Einstein relation $D(x)\Gamma(x)=k_BT$ holds locally and $F(x) = -dV(x)/dx$. In the present case of the numerical modeling the Stokes-Einstein relation holds locally as we have already seen. 

\section{Escape Rates in the presence of Coordinate Dependent Diffusivity}
In this section we will extend the scope of the random walk model to accommodate weakly non equilibrium systems, such as systems modelled by Kramers' escape rate theory. Just as one does in the Kramers' escape rate calculations, we describe a system in which Brownian particles are in a potential well of functional form $U(x)$ (shown in Fig.6) with a local minima at $x=0$ (name of the position $A$ (say)) and a particle sink at $x=45$ (named $B$).  The Smoluchowski equation for the case of a coordinate dependent diffusivity is
\\
\\
$
\frac{\partial P (x,t)}{\partial t}=\frac{\partial}{\partial x} \Big [\frac{P (x,t)}{\Gamma (x)} \frac{\partial U(x)}{\partial x} + \frac{\partial P (x,t) D(x)}{\partial x} \Big]= - \frac{\partial}{\partial x} j(x,t)
$
\\
\\
The steady state probability current $j_{0}$ can be solved for the case of weakly non-equilibrium systems and whose functional form is given by solving
\\
\\
$
\frac{P (x) D(x)}{k_{B}T} \frac{\partial U(x)}{\partial x} + \frac{\partial P (x) D(x)}{\partial x}=-j_{0}
$
\\
\\
where $P(x)$ now refers to the stationary distribution of the system. Writing $P(x)=\frac{1}{D(x)} e ^{\frac{-U(x)}{k_{B}T}} \hat{P} (x)$ gives us
\\
\\
\begin{eqnarray}\nonumber
&&-j_{0}=\frac{e^{\frac{-U(x)}{k_{B}T}}}{k_{B}T} \hat{P} (x)\frac{\partial U(x)}{\partial x} + e^{\frac{-U(x)}{k_{B}T}} \frac{\partial \hat{P} (x)}{\partial x} \\\nonumber &&-\frac{\hat{P} (x)}{k_{B}T} \frac{\partial U(x)}{\partial x} e^{\frac{-U(x)}{k_{B}T}} = e^{\frac{-U(x)}{k_{B}T}} \frac{\partial \hat{P} (x)}{\partial x}
\end{eqnarray}
\\
\\
and on integration we get a closed form for $\hat{P}(x)$ to be 
\\
\\
$
\hat{P}(x) = \hat P(0) - j_{0} \int_{A}^{x} e^{\frac{U(x^{\prime})}{k_{B}T}} dx^{\prime}
$
\\
\\
At $x=A$ the stationary distribution $P(x)$ matches that predicted by the modified Boltzmann Distribution. This gives us
\\
\\
$\hat P(0)= \Big[ \int_{around A} dx^{\prime} \frac{1}{D(x^{\prime})} e^{\frac{-U(x^{\prime})}{k_{B}T}} \Big] ^{-1}$
\\
\\
Near the particle sink $x=B$ , owing to the fact that very few particles manage to escape the potential barrier, one approximately says that $P(x)=0$. This gives us
\begin{equation}
j_{0} = \Big[ \int_{A}^{B} dx^{\prime} e^{\frac{U(x^{\prime})}{k_{B}T}}  \int_{around A} dx^{\prime} \frac{1}{D(x^{\prime})} e^{\frac{-U(x^{\prime})}{k_{B}T}} \Big]^{-1}
\end{equation}
\\

One can contrast this with the expression one were to obtain if one forces the Boltzmann distribution. In this case we have the following Smoluchoski equation
\\
\\
$-j_{0}^{\prime}=D(x) \Big[\frac{P(x)}{k_{B}T} \frac{\partial U(x)}{\partial x} +\frac{\partial P(x)}{\partial x} \Big]$,
\\
\\
and on repeating the calculations mentioned above would obtain a steady state current 
\begin{equation}
j_{0}^{\prime}=\Big[ \int_{A}^{B} \frac{1}{D(x^\prime)}e^{\frac{U(x^{\prime})}{k_{B}T}} dx^{\prime} \int_{around A} dx^{\prime} e^{\frac{-U(x^{\prime})}{k_{B}T}} \Big]^{-1}.
\end{equation}
\\

Note that, there is a change in place of the $1/D(x)$ factor in Eq.8 as compared to that in Eq.7. At present, in the random walk simulation of ours, we know what is the $D(x)$ profile set. Therefore, from a comparison of results of Eq.7 and Eq.8 with that of the simulation we would be able to identify which one works. Note also that, the effect of the $1/D(x)$ in the two different integrals are quite distinct. Where in Eq.7, the $D(x)$ at the position of smaller $U(x)$ will dominate, in Eq.8, the $D(x)$ at a higher $U(x)$ (near the barrier) will have prominence. 
\\ 

The potential of the system is modelled as :
\\
\\
$$
U(x) = \left\{
        \begin{array}{ll}
            2x^{2}-0.001x^{4} & x \geq 0 \\
            2x^{2} & x<0
        \end{array}
    \right.
$$
\\
\\
In what follows the ensemble averages have been set over $10^6$ particles. As shown in Fig.6 this potential well has a minimum at the point $x=0$ and hence close to the bottom of the well, at lower temperatures, the Brownian particles are expected to equilibrate with the bath. Moreover, the escape rates of particles are expected to be very low. This leads to a separation of time scales in the system with the time to equilibrate to the heat-bath near the bottom of the potential well being much smaller than the characteristic time scale for escape of a BP at lower temperatures.  
\\

\begin{figure}[h!]
  \includegraphics[width=60mm]{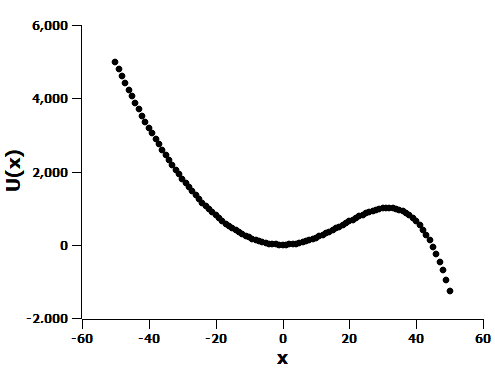}
   \caption{The potential U(x) employed in simulations}
\end{figure}
\hspace*{4mm}
\\ 
\\
\begin{figure}[h!]
  \includegraphics[width=60mm]{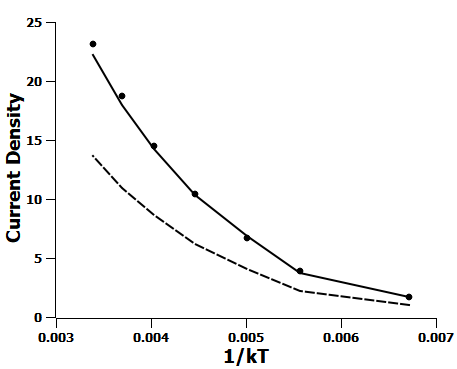}
   \caption{Dependence of ensemble-averaged current density on temperature}
\end{figure}
\hspace*{4mm}
\\ 
\\
\begin{figure}[h!]
  \includegraphics[width=60mm]{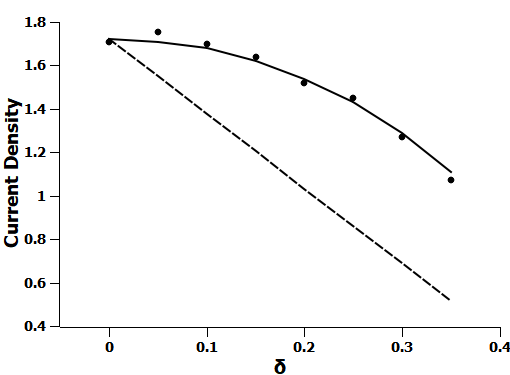}
   \caption{Dependence of ensemble-averaged current density on diffusivity gradient set by $\delta$}
\end{figure}
\hspace*{4mm}
\\

Once again a ramp in the diffusivity is employed by allowing $p_{s}$ to vary spatially in the following manner:
\\
\\
$$
p_{s}(x) = \left\{
        \begin{array}{ll}
            0.2 & x \leq -40 \\
            0.2+\frac{0.6*(x+40)}{80}& -40 \leq x \leq 40 \\
            0.8 & x \geq 40
        \end{array}
    \right.
$$
\\
\\
and $N_{0}$ is set low enough for the particles to equilibrate with the bath. In order to ensure that a steady state is obtained every time a BP reaches the particle sink, which is set at $B=45$, it is immediately taken back to the bottom of the potential well at $x=0$ ($A$). Thus, the total number of $10^6$ particles are always maintained inside the well. The total number of particles that escape the well by each step of the simulation, which we call $N_{left}(t)$, is counted. One step of simulation accounts for the time in which each of $10^6$ particles are attempted to move once. It is observed that for $N_0$ in the admissible range of these simulations, the system took at least 3000 steps of simulation from the initial instant to attain a steady current.
\\

At the steady state the numerically obtained particle distribution near the bottom of the potential well was fitted to the modified Boltzmann Distribution $P(x)D(x)=N exp(\frac{-U(x)}{k_{b}T})$ where $N$ is the normalization constant and the value of  $-\frac{1}{k_{b}T}$ was determined. The steady state current was also numerically obtained by plotting $N_{left}(t)$ versus steps once the system reaches steady state and performing a linear fit. The slope of this plot will be $10^{6}$ times the probability current density which is hence determined. This is done for different values of $N_{0}$ and the numerical results are compared to the values of the current density predicted by Eq.7 and 8. This is shown in Fig.7 in which the solid line represents the current density predicted by Eq.7, the dashed line represents that predicted by Eq.8 and the data points are obtained from simulations.
\\
 
On inspection of Eq.7 one can see that the current density of the system is only affected by the gradient of the diffusivity near the bottom of the potential well. We would like to study the influence of a sharply sloped coordinate dependent diffusivity at the bottom of the potential well on the current density of the system. We will consider the same potential $U(x)$ and set the temperature corresponding to $N_{0}=300$. However we employ a different functional form of $p_{s}$ :
\\
\\
$$
p_{s}(x) = \left\{
        \begin{array}{ll}
            0.5-\delta & x \leq -5 \\
            (0.5+\delta)+(\frac{\delta}{5})(x-5)& -5 \leq x \leq 5 \\
            0.5+\delta & x \geq 5
        \end{array}
    \right.
$$
\\
\\
where delta is varied to obtain different diffusivity gradients. The current density obtained numerically is then compared to those predicted by Eq.7 and 8 by repeating the computational procedure described previously. The results are shown in Fig.8 in which the solid line represents the current density predicted by Eq.7, the dashed line represents that predicted by Eq.8 and the data points are obtained from simulations.
\\

We hence see that the numerical simulations show an excellent agreement with the theoretical considerations of Eq.7 which is most affected by the diffusivity at the bottom of the potential well. Eq.8 that is dominated by the diffusivity at the barrier does not really match the random walk simulation. This is a very interesting physics, that in the coordinate dependent diffusivity the diffusivity at the bottom of the well will play the dominant role in a Kramers' process than the one near the barrier. This is also obvious physics because the Kramers' escape rate process is dominated by the equilibrium distribution. Thus, this observation could be the candidate for major experimental validation of which one is the equilibrium distribution for a confined BP in the presence of coordinate dependent diffusivity. 

\section{discussion}
In this paper, we have established a numerical method based on simple random walk on lattice to investigate a BP with coordinate dependent diffusivity in a constant temperature environment. This Brownian motion, by construction, is such that the Stokes-Einstein relation holds locally i.e. $D(x)\Gamma(x) =k_BT$. The temperature of the system can be tuned by a parameter which determines how much up the potential barrier the BP is allowed to move. This simple model captures an ideal situation for the equilibrium of a BP with coordinate dependent diffusivity at a constant temperature with a heat-bath.
\\

The result shows excellent agreement with the theoretical prediction that the equilibrium distribution of such a BP under confinement will be a modified Boltzmann distribution with a diffusivity dependent amplitude. This is a very important result which actually encodes in the equilibrium distribution the sources of both the inhomogeneity of space, namely, the conservative force and the coordinate dependent diffusivity. In the presence of the local validity of the Stokes-Einstein relation, the damping is slaved by the diffusivity and that is well captured by our model system. The present computational model can work for any confining potential. We have considered here a weakly non-euilibrium situation of particle leaking over a barrier as well. The equilibrium distribution works in such weakly non-equilibrium cases. We show that the agreement of the results of simulation with the modified Boltzmann distribution is excellent.
\\

In the analytic derivation of the Kramers' rate of transition over a barrier we have seen that the Boltzmann distribution shows a dependence of the process on the diffusivity at the barrier whereas the modified Boltzmann distribution captures that such rate processes will be dominated by the diffusivity at the well minimum. The latter is a much more acceptable scenario on physical basis that in such processes the BP spends more time at the potential well than on top of the barrier. The result obtained from modified Boltzmann distribution is corroborated by the simulation using random walk. This result, when experimentally verified can alleviate many confusions as to how to deal with such systems of Brownian motion in the presence of coordinate dependent diffusion.
\\

We have presented in this work a very simple method of simulation on lattice which can in general be employed to any form of conservative force field at a constant temperature with a coordinate dependent diffusivity. However, despite being a very simple method, this numerical analysis has not yet been used much in the place of Langevin dynamics of systems with coordinate dependent diffusivity. The present paper establishes the reliability of this simple method for use in such systems where the form of the Langevin dynamics is controversial due to many issues related to many conventions. One can explore this method of modeling with random walk for biological processes where coordinate dependence of the diffusivity is needed at a constant temperature, for example, protein folding on a lattice.

\end{document}